\documentclass[%
 aip,
 amsmath,amssymb,
preprint,
]{revtex4-1}

\usepackage[dvipdfmx]{graphicx}
\usepackage{dcolumn}% Align table columns on decimal point
\usepackage{bm}% bold math
\usepackage[utf8]{inputenc}
\usepackage[T1]{fontenc}
\usepackage{mathptmx}
\usepackage{etoolbox}
\usepackage{color}

\newcommand {\ECA}{EuCd$_2$As$_2${}}

\newcommand {\TN}{$T_\mathrm{N}${}}

\newcommand {\Ms}{$M_\mathrm{s}${}}

\newcommand {\rhoxx}{$\rho _{xx}${}}
\newcommand {\rhoyx}{$\rho _{yx}${}}

\makeatletter
\def\@email#1#2{%
 \endgroup
 \patchcmd{\titleblock@produce}
  {\frontmatter@RRAPformat}
  {\frontmatter@RRAPformat{\produce@RRAP{*#1\href{mailto:#2}{#2}}}\frontmatter@RRAPformat}
  {}{}
}%
\makeatother
\begin{document}

\preprint{AIP/123-QED}

\title{Intrinsic insulating transport characteristics in low-carrier density EuCd$_2$As$_2$ films}
% Force line breaks with \\
\author{Shinichi Nishihaya}
\affiliation{ 
Department of Physics, Tokyo Institute of Technology, Tokyo, 152-8551, Japan%\\This line break forced with \textbackslash\textbackslash
}
\author{Ayano Nakamura}%
\affiliation{ 
Department of Physics, Tokyo Institute of Technology, Tokyo, 152-8551, Japan%\\This line break forced with \textbackslash\textbackslash
}

\author{Mizuki Ohno}
\affiliation{ 
Department of Applied Physics and Quantum-Phase Electronics Center (QPEC), the University of Tokyo, Tokyo 113-8656, Japan%\\This line break forced with \textbackslash\textbackslash
}%

\author{Markus Kriener}
\affiliation{ 
RIKEN Center for Emergent Matter Science (CEMS), Wako 351-0198, Japan%\\This line break forced with \textbackslash\textbackslash
}%

\author{Yuto Watanabe}
\affiliation{ 
Department of Physics, Tokyo Institute of Technology, Tokyo, 152-8551, Japan%\\This line break forced with \textbackslash\textbackslash
}

\author{Masashi Kawasaki}
\affiliation{ 
Department of Applied Physics and Quantum-Phase Electronics Center (QPEC), the University of Tokyo, Tokyo 113-8656, Japan%\\This line break forced with \textbackslash\textbackslash
}%
\affiliation{ 
RIKEN Center for Emergent Matter Science (CEMS), Wako 351-0198, Japan%\\This line break forced with \textbackslash\textbackslash
}%

\author{Masaki Uchida}
\altaffiliation{Electronic mail: m.uchida@phys.titech.ac.jp}
\affiliation{ 
Department of Physics, Tokyo Institute of Technology, Tokyo, 152-8551, Japan%\\This line break forced with \textbackslash\textbackslash
}


\begin{abstract}
Searching for an ideal magnetic Weyl semimetal hosting only a single pair of Weyl points has been a focal point for systematic clarification of its unique magnetotransport derived from the interplay between topology and magnetization. Among the candidates, triangular-lattice antiferromagnet \ECA\ has been attracting special attention due to the prediction of the ideal Weyl semimetal phase in the ferromagnetic state, however, transport properties of low-carrier density samples have remained elusive. Here we report molecular beam epitaxy growth of \ECA\ films, achieving low-hole density in the range of $10^{15}$-$10^{16}$ cm$^{-3}$ at low temperature. Transport measurements of such low-carrier density films reveal an insulating behavior with an activation gap of about 200 meV, which persists even in the field-induced ferromagnetic state. Our work provides an important experimental clue that \ECA\ is intrinsically insulating, contrary to the previous prediction. 
\end{abstract}

\maketitle
Magnetic Weyl semimetals (WSMs) with three-dimensional band degenerate points (Weyl points) formed between the time-reversal-symmetry-broken bands, has attracted research interest, especially as a platform hosting a rich interplay between topology and magnetism\cite{Review,mWSM}. In addition to the magnetotransport phenomena derived from the giant Berry phase of Weyl points, such as large anomalous Hall/Nernst effects\cite{mWSM,Mn3Sn_AHE,CoMnGa_AHE,CoSnS_AHE,ECS} and chiral anomaly\cite{Anomaly1,Anomaly2,Anomaly3}, unique coupling between charge/spin transport and magnetization potentially leads to novel spintronics functionalities\cite{Spintronics1,Spintronics2,Spintronics3}. The Weyl points position in the momentum space of magnetic WSMs is generally subject to symmetry modulation under different magnetic ordering states, which also provides a useful tunability of the Weyl transport by external means such as magnetic field and strain\cite{WPmove,WPmove2}. 

From the symmetry point of view, the time-reversal symmetry-broken WSMs allow the formation of a single pair of Weyl points with opposite chirality, in contrast to the inversion-symmetry-broken WSMs, which possess at least two pairs of Weyl points at the same energy\cite{mWSM}. However, the actual band structure in most of the magnetic WSM materials experimentally-established so far is more complicated by the formation of multiple pairs of Weyl points at different energies and also by the occupation of non-Weyl bands\cite{MnSn_ARPES,CoSnS_ARPES,CoMnGa_ARPES}. A more simplified band structure with ideally a single pair of Weyl points near the Fermi energy is desired to systematically evaluate and control exotic magnetotransport derived from the Weyl points. 
  
In this respect, triangular-lattice antiferromagnet \ECA , which has been proposed to exhibit a WSM phase with only a single pair of Weyl points in the ferromagnetic state\cite{Theory2}, has gathered attention as potentially the most ideal material candidate. Since the prediction of an antiferromagnetic Dirac semimetal phase\cite{Theory1} and a field-induced ferromagnetic WSM phase\cite{Theory2} in \ECA , various experimental efforts have been made to elucidate its electronic and magnetic features based on bulk crystals, including angle-resolved photo-emission microscopy (ARPES)\cite{ARPES}, neutron diffraction\cite{ND}, and transport measurements reporting quantum oscillations\cite{QO} and anomalous Hall/Nernst effects\cite{nonAHE,ANE}. In particular, these transport observations have revealed interesting  properties of \ECA , which are potentially associated with the predicted WSM band structure, however, their limitation lies in the high carrier density of the bulk samples, hindering the direct evaluation of the gapless nature of the low energy state. Moreover, the non-trivial picture of the band structure of \ECA\ has been recently challenged from both the experimental and theoretical sides; A pump-probe spectroscopy study on \ECA\ bulk crystals has reported the observation of a 0.8 eV band gap\cite{Semicon}, and a theoretical study points out the possibility that the non-trivial band inversion energy tends to be overestimated and that \ECA\ is actually a trivial insulator\cite{Overestimation}. From these aspects, it is of crucial importance to evaluate transport properties of the low-carrier density samples.

Here we report growth of low-carrier-density film samples of \ECA\ by molecular beam epitaxy (MBE). Owing to the growth under Cd and As overpressure, the hole density of these \ECA\ films is highly suppressed to $10^{15}$-$10^{16}$ cm$^{-3}$. Importantly, such low-carrier density \ECA\ exhibit a clear activation-type insulating behavior with high resistivity even in the field-induced ferromagnetic state, in contrast to the gapless semimetal phase predicted for \ECA . 

%%%%%%%%%%%%%%%%%%%%%%%%%%%%%%%%%%%%%%%%%%%%%%%%%%%%%%%%%%%%%%%%%%%%%%%%%%%%%%%%%%%%%%%%%%%%%%%%%%%%%%%%%%%

\ECA\ films were grown by MBE technique on the CdTe(111)A surface which has the same three-fold symmetry and relatively small lattice mismatch ($+3.2$\%) with respect to \ECA (001), as compared in Figs. 1(a) and 1(b). Prior to the growth of \ECA , CdTe substrates were etched with 0.01\% Br$_2$-methanol for 5 min to remove the native oxide layer, and then annealed at 520$^{\circ}$C under Cd overpressure inside the MBE chamber (Epiquest RC1100) to achieve a smoother surface\cite{ECS,CdTeEtching}. Figure 1(d) shows a RHEED pattern taken on the CdTe(111)A surface after the thermal annealing, showing the diffraction pattern with the partially Cd-deficient ($2\times 2$) surface reconstruction\cite{CdTeSurface}. 

\ECA\ films were grown by co-evaporation of the elemental sources; Eu (3N, Nippon Yttrium Co. Ltd.) and Cd (6N, Osaka Asahi Co.) using a conventional Knudsen cell, and As (7N5, Furukawa Co.) in the form of As$_2$ using a MBE-Komponenten valved-cracker cell. The substrate temperature was initially set to 200$^{\circ}$C for the nucleation of the first 5 nm seed layer in an effort to suppress the interfacial reaction between \ECA\ and CdTe, and then was raised to higher temperatures (205-240$^{\circ}$C) for the rest of the growth. The film thickness was designed to be 30-50 nm. Due to the highly volatile nature of Cd and As compared to Eu, the growth was performed under Cd- and As-rich conditions, with the ratios between the beam equivalent pressure of the sources of Cd/Eu > 400 and As/Eu = 4-5. This condition is critical to achieve highly-crystalline \ECA\ films with low-carrier density. 

%%%%%%%%%%%%%%%%%%%%%%%%%%%%%%%%%%%%%%%%%%%%%%%%%%%%%%%%%%%%%%%%%%%%%%%%%%%%%%%%%%%%%%%%%%%%%%%%%%%%%%%%%%%

Figure 1(c) presents RHEED diffraction patterns taken on the \ECA\ (001) surface, and Fig. 1(e) shows $\theta$-2$\theta$ x-ray diffraction (XRD) results for a 35 nm thick \ECA\ film. The diffraction peaks from the \ECA(001) plane oriented to the surface normal are clearly observed. We have also confirmed that EuTe tends to form at the interface between \ECA\ and CdTe, which is possibly induced by the fact that the nucleation of \ECA\ initiates on the Cd-poor surface of the CdTe (111)A substrates subject to high-temperature annealing (see also the supplementary material). The reciprocal space map presented in Fig. 1(f) reveals the in-plane stacking relation between \ECA\ and CdTe, where the $a$-axis of \ECA\ is aligned to the [210] direction of CdTe. We also emphasize that \ECA\ films show a three-fold symmetry as expected from its crystal structure, and 60$^\circ$ rotated domains are negligibly minimal as shown in the $\varphi$ scan presented in Fig. 1(g). For the current thickness range (30-50 nm), \ECA\ films are fully relaxed on the CdTe substrate (see the supplementary material for details).

%%%%%%%%%%%%%%%%%%%%%%%%%%%%%%%%%%%%%%%%%%%%%%%%%%%%%%%%%%%%%%%%%%%%%%%%%%%%%%%%%%%%%%%%%%%%%%%%%%%%%%%%%%%%

\ECA\ has been known to exhibit different magnetic ground states sensitively depending on the off-stoichiometry\cite{ANE,FM2020} and the lattice parameter changes by external pressure\cite{Pressure1,Pressure2}. Most of the bulk single crystals at ambient pressure show an A-type antiferromagnetic ordering, where the Eu$^{2+}$ moments order ferromagnetically within each Eu layer and antiferromagnetically between the adjacent Eu layers\cite{ND,QO,nonAHE,Pressure1,Pressure2,AFM2020,AFM2021,AFM2022}. It has been reported that bulk crystals with a subtle Eu deficiency exhibit a ferromagnetic ground state with a Curie temperature around 26 K\cite{ANE,FM2020}. Figure 2(a) shows temperature dependence of magnetization of the \ECA\ film measured with an out-of-plane magnetic field 0.1 T applied along the $c$-axis. A clear kink structure corresponding to an antiferromagnetic transition is observed, and no signatures of other magnetic ordering are confirmed. The N\'{e}el temperature \TN\ varies slightly according to the lattice parameters in the range of 9.1-9.4 K (see also the supplementary material for the sample dependence). The magnetization curves measured at different temperatures are presented in Fig. 2(b). The magnetization increases monotonically as increasing the out-of-plane field, and saturates above 2 T. This can be understood simply by the magnetization process where the antiferromagnetically-ordered in-plane moments are gradually canted to the out-of-plane field direction without any additional magnetic transitions. \textcolor{black}{We note that EuTe has been known to show a antiferromagnetic order around 9.6 K\cite{EuTe_mag}, which is close to \TN\ of \ECA . However, the magnetic contribution from the interfacial EuTe layer should be small considering its relative volume to that of the \ECA\ layer judged from the x-ray diffraction intensity. The saturated behavior of the magnetic moments around 2 T for the \ECA\ film is also distinguished from the magnetic response of EuTe which typically exhibits a higher saturation field around 8 T\cite{EuTe_mag}.}

%%%%%%%%%%%%%%%%%%%%%%%%%%%%%%%%%%%%%%%%%%%%%%%%%%%%%%%%%%%%%%%%%%%%%%%%%%%%%%%%%%%%%%%%%%%%%%%%%%%%%%%%%%%    

Next we present the transport properties of the \ECA\ films. Figure 3(a) shows temperature dependence of resistivity \rhoxx\ measured for several representative samples (Films A, B, and C). While the bulk crystals reported in previous studies show metallic conduction for a wide temperature range\cite{ARPES,ND,QO,nonAHE,Semicon,AFM2021}, \ECA\ films exhibit an insulating behavior below the room temperature. \rhoxx\ exceeds 10$^3$ $\Omega$cm at 150 K for Film C, and the sheet resistance reaches an immeasurably high value (over $10^7$ $\Omega$) at lower temperatures for all the films. Importantly, such an insulating behavior with high resistivity remains even under 9 T cooling, contrary to the theoretical prediction that a gapless WSM phase is realized in the field-induced ferromagnetic state\cite{Theory2}. 

To clarify the origin of the insulating behavior, Hall measurements have been performed at high temperatures. Figures 3(b) and 3(c) summarize field dependence of Hall resistivity \rhoyx\ for Films A and B, respectively. The \rhoyx\ curves can be well fitted by a two-carrier transport model with dominant low-mobility p-type conduction and higher-mobility n-type conduction components. \textcolor{black}{Here, the origin of the minority n-type conduction, which is below 10\% of the overall sheet conductance, is not clear and further study is required. As discussed in the supplementary material, the most likely origin is the interface between the \ECA\ film and the CdTe substrate, either due to formation of the EuTe layer which is known to be an easily electron-doped semiconductor\cite{EuTe}, or off-stoichiometry of the \ECA\ layer near the interface. On the other hand, the dominant p-type conduction is attributed to the intrinsic transport contribution of the \ECA\ film, which is consistent with the easily hole-doped nature of \ECA\ reported for bulk crystals in the previous studies\cite{ARPES,ND,QO,nonAHE,Semicon,AFM2021}.} Figure 3(d) presents the relation between \rhoxx\ and hole density $p$ of the \ECA\ films plotted together with the bulk crystal data\cite{ND,QO,nonAHE,AFM2021}. The hole density of the \ECA\ films is in the range of $10^{15}$-$10^{16}$ cm$^{-3}$ at low temperatures, which is much lower than the previously-reported values for the bulk crystals. The hole density exhibits an activation-type behavior and is suppressed by several orders of magnitudes with decreasing temperature, strongly indicating the presence of an energy gap in the band structure. According to the conventional Arrhenius plot, the activation energy is estimated to be 180 meV for Film A and 210 meV for Film B. We note that the hole mobility (about $20$ cm$^2$/Vs for Film A, see the supplementary material) does not show significant change during the cooling, ruling out the carrier localization scenario for the observed insulating behavior.

%%%%%%%%%%%%%%%%%%%%%%%%%%%%%%%%%%%%%%%%%%%%%%%%%%%%%%%%%%%%%%%%%%%%%%%%%%%%%%%%%%%%%%%%%%%%%%%%%%%%%%%%%%%%

While the previous transport studies based on high-hole-density bulk crystals have reported metallic conduction\cite{ARPES,ND,QO,nonAHE,AFM2021}, our low-carrier-density films reveal an intrinsic insulating behavior.  We also emphasize that the observed insulating behavior with an about 200 meV activation gap is not induced by the confinement effect. According to the ARPES measurements\cite{ARPES}, the Fermi velocity of the bulk bands forming the hole pocket at the zone center is $6 \times 10^{4}$ m/s, and a simple estimation of the confinement gap for 35 nm thickness gives only 20 meV. In an effort to measure the low-temperature transport, we have also fabricated a top-gate transistor device on a \ECA\ film to electrostatically accumulate carriers. As discussed in detail in the supplementary material, the \ECA\ film remains to be highly-insulating at low temperatures even under the application of a hole accumulating gate voltage, which further underlines the presence of a large enough energy gap to prevent the carrier accumulation by electrostatic band bending. 

In summary, we have grown epitaxial films of \ECA\ which has been long thought to be an ideal magnetic WSM candidate. Owing to the growth with well-calibrated flux ratios, we have achieved low-carrier density \ECA\ films. Through the transport measurements, we have revealed that the \ECA\ films exhibit an activation-type insulating behavior, which evidences the presence of an intrinsic band gap greater than the activation energy. Our work provides an important experimental clue for understanding the controversial electronic band structure of \ECA . 

\section*{Supplementary Material}
See the supplementary material for (1) sample dependence of the structural and magnetic properties of \ECA\ films, (2) two-carrier fitting, and (3) electrostatic gating on the \ECA\ film.

\begin{acknowledgments}
\textcolor{black}{This work was supported by JST FOREST Program Grant No. JPMJFR202N, Japan, by Grant-in-Aids for Scientific Research JP21H01804, JP22H04471, JP22H04501, JP22H04958, JP22K18967, JP22K20353, and JP23K13666 from MEXT, Japan, and by 2022 Yoshinori Ohsumi Fund for Fundamental Research, Japan.}
\end{acknowledgments}

\section*{Conflicts of Interest}
The authors have no conflicts to disclose.
\section*{Author Contributions}

\textbf{Shinichi Nishihaya}: Conceptualization (equal); Data curation (lead); Formal analysis (lead); Investigation (lead); Methodology (equal); Visualization (lead); Writing - original draft preparation (lead). Funding acquisition (equal)
\textbf{Ayano Nakamura}: Formal analysis (supporting); Investigation (equal); Methodology (equal); Writing - review \& editing (equal).
\textbf{Mizuki Ohno}: Formal analysis (supporting); Investigation (equal); Methodology (equal); Writing - review \& editing (equal).
\textbf{Markus Kriener}: Formal analysis (supporting); Investigation (supporting); Methodology (equal); Writing - review \& editing (equal).
\textbf{Yuto Watanabe}: Formal analysis (supporting); Investigation (supporting); Methodology (equal); Writing - review \& editing (equal).
\textbf{Masashi Kawasaki}: Formal analysis (supporting); Investigation (supporting); Methodology (equal); Writing - review \& editing (equal). Funding acquisition (lead).
\textbf{Masaki Uchida}: Conceptualization (equal); Funding acquisition (lead); Methodology (equal); Project administration (lead); Supervision (lead); Writing - review \& editing (equal).

\section*{Data Availability}
The data that supports the findings of this study are available
from the corresponding author upon reasonable request.

%%%%%%%%%%%%%%%%%%%%%%%%%%%%%%%%%%%%%%%%%%%%%%%%%%%%%%%%%%%%%%%%%%%%%%%%%%%%%%%%%%%%%%%%%%%%%%%%%%%%%%%%%%% 

%\section*{Reference}
%\bibliography{PtMnSb_ref}% Produces the bibliography via BibTeX.

\newpage 

%%%%%%%%%%%%%%%%%%%%%%%%%%%%%%%%%%%%%%%%%%%%%%%%%%%%%%%%%%%%%%%%%%%%%%%%%%%%%%%%%%%%%%%%%%%%%%%%%%%%%%%%%%% 

\begin{figure*}
%\begin{figure}
\begin{center}
\includegraphics[width=16.5cm]{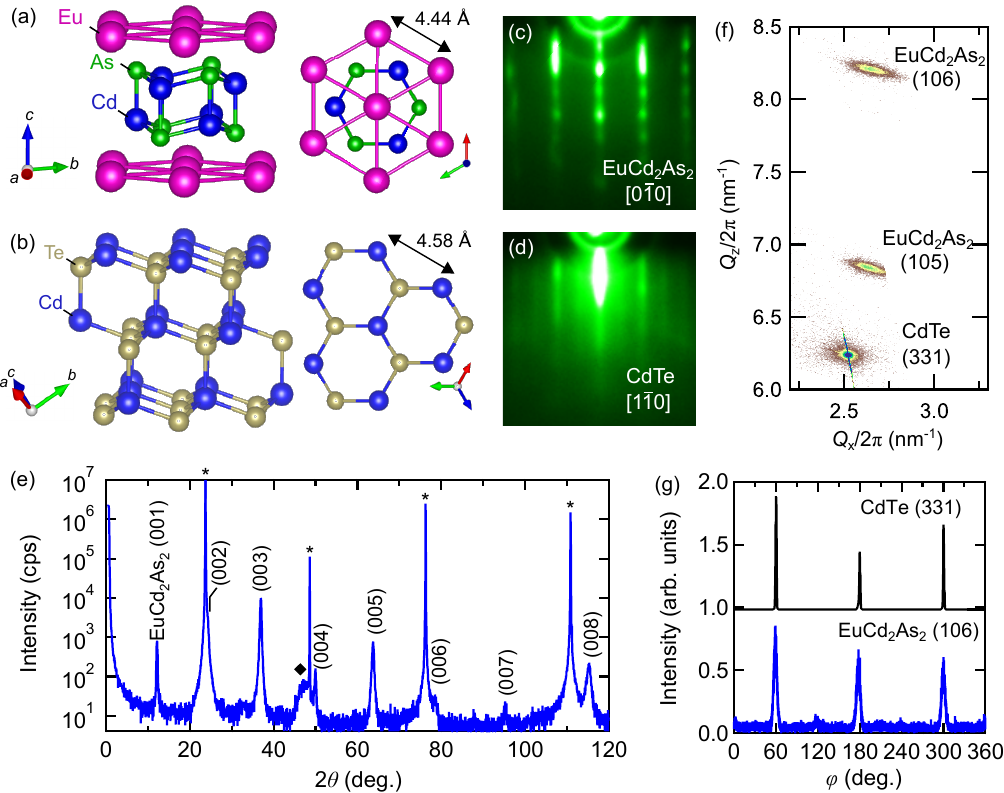}
\caption{
Structural characterization of \ECA\ films grown on the CdTe (111)A surface. Crystal structure of (a) \ECA\ and (b) CdTe. RHEED pattern for (c) the \ECA\ film and (d) the CdTe substrate from the CdTe$\left[ 1\bar{1}0\right] $ azimuth (parallel to \ECA$\left[ 0\bar{1}0 \right]$). (e) $\theta$-2$\theta$ XRD scan on a 35 nm thick \ECA\ film, showing the $c$-axis of \ECA\ is oriented to the surface normal. The asterisk mark denotes the diffraction peak from the CdTe(111) substrate, and the solid diamond the EuTe impurity phase formed at the interface. (f) Reciprocal space map collected around the CdTe and \ECA\ Bragg peaks. (g) Three-fold symmetry observed in polar XRD measurements for the CdTe(331) and \ECA  (106) peaks, as expected from their $C_{3}$ rotation-symmetric crystal structure shown in (a) and (b).
}
\label{fig1}
\end{center}
\end{figure*}
%\end{figure}

\begin{figure}
%\begin{figure*}
\begin{center}
\includegraphics[width=7cm]{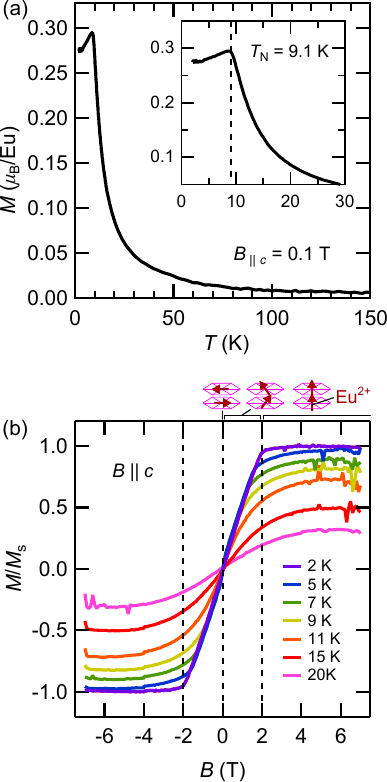}
\caption{
Magnetization properties of \ECA\ films. (a) Temperature dependence of Magnetization measured on a representative film under out-of-plane magnetic field $B_{\parallel c}$ = 0.1 T. A clear kink structure corresponding to the antiferromagnetic transition appears at \TN\ = 9.1 K. (b) Magnetization curves measured with out-of-plane magnetic fields at different temperatures. Data are normalized with respect to the saturation moment \Ms\ at 2 K. The spin canting process is schematically shown on the top.
}
\label{fig2}
\end{center}
\end{figure}
%\end{figure*}

%\begin{figure}
\begin{figure*}
\begin{center}
\includegraphics[width=17cm]{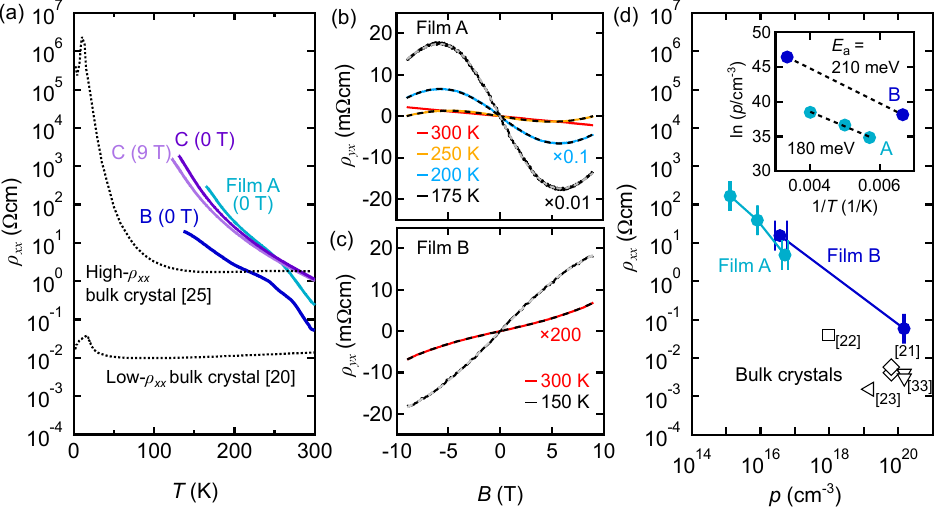}
\caption{
Transport properties of the low-carrier-density \ECA\ films. (a) Temperature dependence of resistivity \rhoxx\ of different \ECA\ films (A, B, and C). The data of bulk crystals reported in the previous studies\cite{ARPES,Semicon} are also shown for comparison. The \ECA\ films exhibit an insulating behavior from high temperatures, which remains the same even under 9 T cooling (Film C). Hall resistivity \rhoyx\ measured at different temperatures for (b) Film A, and (c) Film B. The overlayed broken lines are the results of two-carrier model fitting. (d) Relation between \rhoxx\ and hole density $p$, indicating an activation-type behavior in the \ECA\ films. The data of bulk crystals\cite{ND,QO,nonAHE,AFM2021} are also shown for comparison. The inset shows the Arrhenius plot for estimating the activation energy $E_{\mathrm{a}}$.
}
\label{fig3}
\end{center}
%\end{figure}
\end{figure*}

%%%%%%%%%%%%%%%%%%%%%%%%%%%%%%%%%%%%%%%%%%%%%%%%%%%%%%%%%%%%%%%%%%%%%%%%%%%%%%%%%%%%%%%%%%%%%%%%%%%%%%%%%%%%

\end{document}